\documentclass[reprint,aps,amsmath,amssymb,amsfonts,showkeys]{revtex4-2}
\usepackage[T1]{fontenc}
\usepackage[pdftex]{graphicx,xcolor}
\usepackage{dcolumn}
\usepackage{times}
\usepackage{multirow}
\usepackage{physics}
\usepackage{siunitx}
\usepackage{here}
\AtBeginDocument{\RenewCommandCopy\qty\SI}

\usepackage{xspace}
\usepackage{ulem}
\newcommand{\cco}{Ca$_3$Co$_4$O$_9$\xspace}
\newcommand{\sn}{Ca$_3$Co$_{4-x}$Sn$_x$O$_9$\xspace}

\usepackage[pdftex,bookmarks=false,bookmarksnumbered=false,colorlinks=true,linkcolor=black,citecolor=blue,filecolor=black,urlcolor=blue]{hyperref}

\begin{document}
\title{Impurity-induced spin density wave in the thermoelectric layered cobaltite [Ca$_2$CoO$_3$]$_{0.62}$[CoO$_2$]}

\author{Motoya~Takenaka$^1$}
\email{6224522@ed.tus.ac.jp}
\author{Shogo~Yoshida$^1$}
\email{s_yoshida@rs.tus.ac.jp}
\author{Yoshiki~J.~Sato$^1$}
\thanks{Present address: Graduate School of Science and Engineering, Saitama University, Saitama 338-8570, Japan}
\author{Ryuji~Okazaki$^1$}
\affiliation{$^1$Department of Physics and Astronomy, Tokyo University of Science, Noda 278-8510, Japan}

\begin{abstract}
We investigate the Sn-substitution effect on the thermoelectric transport properties of the layered cobaltite [Ca$_2$CoO$_3$]$_{0.62}$[CoO$_2$] single crystals, 
which exhibit a non-monotonic temperature variation of the electrical resistivity and the Seebeck coefficient owing to the complex electronic and magnetic states.
We find that the onset temperature of the short-range spin-density-wave (SDW) formation increases with the substituted Sn content, 
indicating the impurity-induced stabilization of the SDW order, reminiscent of the disorder/impurity-induced spin order in the cuprate superconductors.
The Seebeck coefficient is well related to such impurity effects, as it is slightly enhanced below the onset temperatures, implying a decrease in the carrier concentration due to a pseudo-gap formation associated with the SDW ordering.
We discuss the site-dependent substitution effects including earlier studies, and 
suggest that substitution to the conducting CoO$_2$ layers is essential to increase the onset temperature, 
consistent with the impurity-induced SDW picture realized in the conducting layers with the cylindrical Fermi surface.

\end{abstract}

\maketitle

\section{introduction}

Thermoelectric materials, which can directly convert heat into electricity and \textit{vice versa} through the Seebeck and Peltier effects, have garnered significant attention owing to their potential applications in energy harvesting and waste heat recovery \cite{Hebert2016,He2017,Shi2020,Yan2022}. 
In particular, 
since the discovery of high thermoelectric performance in Na$_x$CoO$_2$ \cite{Terasaki1997},
thermoelectric oxides have attracted interest due to their excellent thermal stability, environmental friendliness, and abundance of raw materials, making them promising candidates for sustainable energy applications \cite{Yin2017, Ge2024}. 

Among these thermoelectric oxides, the layered cobaltite [Ca$_2$CoO$_3$]$_{p}$[CoO$_2$] ($p\approx 0.62$), also known as the approximate formula \cco, has been extensively studied due to its high thermoelectric performance as well as the unique structural and electronic properties \cite{Masset2000,Miyazaki2002}. \cco consists of alternately stacked CdI$_2$-type conductive CoO$_2$ layers and rocksalt-type Ca$_2$CoO$_3$ block layers, which form two subsystems with different lattice parameters on the $b$ axis \cite{Masset2000,Miyazaki2002}. 
This unique layer arrangement
is referred to as a misfit structure, as is also seen in the transition-metal dichalcogenides \cite{Meerschaut1995,10.1063/5.0101429}. 
The complex electronic and magnetic states have been observed in \cco;
Near room temperature, an incoherent hopping transport of the correlated holes is realized near room temperature, and 
a Fermi-liquid (FL) state with a typical temperature variation of the resistivity $\rho(T)=\rho_0+AT^2$ is then observed below a crossover temperature $T^*\sim 140$~K \cite{PhysRevB.71.233108}.
Subsequently, a short-range order of an incommensurate spin density wave (SDW) appears below the onset temperature $T_{\rm SDW}^{\rm on} \sim 100$~K \cite{Sugiyama2002,PhysRevB.68.134423,PhysRevB.96.035126,PhysRevB.102.094428}, below which the resistivity increases with cooling owing to a pseudo-gap formation associated with the SDW order \cite{Heieh2014}, while the localization effect may also be involved.
The incommensurate SDW order becomes long-range order below $T_{\rm SDW} \sim 30$~K, 
while the magnetic susceptibility shows no apparent anomaly at $T_{\rm SDW}$ \cite{Sugiyama2002}. 
At lower temperatures, a ferrimagnetism develops below $T_{\rm FR}\sim 19$~K.

External perturbations such as physical pressure and elemental substitutions are effective ways to modify such complicated electronic states in \cco,
as might be expected from a rich electronic phase diagram in Na$_x$CoO$_2$ \cite{PhysRevLett.92.247001}.
Indeed, the strong correlation effect is well controlled by applying physical pressure \cite{PhysRevB.71.233108}.
On the other hand, various elemental substitution studies have been performed \cite{Wang2002,Wang2010_3,Pinitsoontorn2012,Fang2024}, 
but most of these studies have been conducted on the polycrystalline samples, aimed at improving thermoelectric performance.
In contrast, 
unusual transport properties have been revealed in several substitution studies using single-crystalline samples; 
In the single-crystalline Bi- and Cu-substituted samples \cite{Mikami2006,Huang2012},
both the electrical conductivity and the Seebeck coefficient increase with elemental substitutions,
in sharp contrast to their conventional trade-off relationship as seen in the 
isovalent substitution systems \cite{Ikeda2016,Saito2017}.
In Ca$_{3-x}$Bi$_x$Co$_4$O$_9$ single crystals, for instance,
Bi$^{3+}$ substitution for 
the Ca$^{2+}$ sites in the Ca$_2$CoO$_3$ layers acts as the electron doping into the conductive CoO$_2$ layer,
leading to decreased hole concentration to explain the increase of the Seebeck coefficient \cite{Mikami2006}.
However, 
this contradicts the observation of increased electrical conductivity, indicating an unusual enhancement of the carrier mobility by such substitutions.
Indeed, such a situation is different from a conventional electron-doping case in the single-crystalline Ca$_3$Co$_{4-x}$Ti$_x$O$_9$,
in which the Seebeck coefficient increases but the electrical conductivity decreases with increasing Ti$^{4+}$ content \cite{Zhao2006}.

In this paper, we have performed the transport study of Sn-substituted \cco single crystals,
electron-doped systems with non-magnetic Sn$^{4+}$ ions
as in the cases of Bi and Ti substitutions.
Near room temperature, both the Seebeck coefficient and the electrical resistivity increase with the Sn content, 
as expected from the decrease of the hole concentration.
Interestingly,
in the Sn-substituted case,
the onset temperature of the short-range SDW formation $T_{\rm SDW}^{\rm on}$ monotonically increases with the Sn content, 
in sharp contrast to the Ti-substituted study where the transition temperature is essentially constant with the Ti content \cite{Zhao2006}.
This result thus
indicates an impurity-induced stabilization of the SDW order in \cco,
where the dynamical spin fluctuations are frozen out by the impurity,
as is discussed in the layered cuprate superconductors \cite{PhysRevLett.99.147002,PhysRevLett.105.147002,Schmid_2013,JANG2023169164}.
The different substitution effects on the onset temperature between the Sn$^{4+}$ and the Ti$^{4+}$ may originate from site-dependent substitutions, since the Co sites in the conductive layers are substituted by the Sn atoms, directly affecting the SDW order.
We also discuss the effect of the pseudogap opening associated with the SDW order on the Seebeck coefficients in \cco.

\section{Methods}

Single crystals of \sn were prepared using a modified SrCl$_2$-flux technique as reported by Shikano $et$ $al$ \cite{Shikano2003}. 
Powders of CaCO$_3$(99.9\%), Co$_3$O$_4$(99.9\%), and SnO$_2$(99.9\%) were mixed in the stoichiometric ratio, and SrCl$_2$(99.9\%) powder was added as the flux with a molar ratio of 1:4.5.
The mixed powders were calcined in air at 1173 K for 3 h, and then the furnace was cooled down to 400 K at a rate of 6 K/h. 
The grown crystals were washed in hot water to remove the SrCl$_2$ flux. 
After the crystal growth, we annealed the crystals in oxygen atmosphere at 973 K for 24 h.
The obtained crystals were thin platelets with typical dimensions of 2 $\times$ 2 $\times$ 0.1 mm$^3$. 
We also prepared the polycrystalline samples of \sn using the solid-state reaction method for the powder X-ray diffraction (XRD) measurements. 
Powders of CaCO$_3$(99.9\%), Co$_3$O$_4$(99.9\%), and SnO$_2$(99.9\%) were mixed in a stoichiometric ratio and calcined two times in air at 1173 K for 24 h with intermediate grindings.

The XRD measurement was performed with Cu K$\alpha$ radiation in a $\theta$-$2\theta$ scan mode. 
The chemical compositions of Ca, Co, Sn, and O were determined by wavelength-dispersive X-ray spectroscopy (WDS) equipped with the Electron Probe Micro analyzer (EPMA, JEOL Ltd., JXA-8100). 
Ca$_3$(VO$_4$)$_2$, CoO, Sn, and SrTiO$_3$ were used as the standard samples purchased from JEOL Ltd. 
Impurities such as Sr were not detected within the experimental resolution.
The $a$-axis electrical resistivity was measured using a standard four-terminal method with a low excitation current of $I=50$ $\mu$A provided by a Keitheley 6221 current source \cite{Sakabayashi2021}. 
The sample voltage was measured with a Keithley 2182A nanovoltmeter. 
The $a$-axis thermopower was measured using a steady-state technique using a manganin-constantan differential thermocouple in a closed-cycle refrigerator \cite{PhysRevB.105.184507,PhysRevB.108.235115,PRXEnergy.3.043007}.
A typical temperature gradient of 0.2 K/mm, which is adjusted along with the bath temperature, was applied along the $a$-axis direction using a resistive heater and the distance between the thermocouple contacts is about 1 mm. 
The thermoelectric voltage from the wire leads was subtracted.
The magnetization of the single crystals was measured using a superconducting quantum interference device (SQUID) measurement system (MPMS).
Since the single crystal is very thin, so we used several dozen crystals.
The magnetic field $\mu_0H=1$~T was applied to the $c$-axis direction and the magnetization $M$ was measured under zero-field cooling condition to obtain the magnetic susceptibility $\chi=M/H$.

\section{Results and Discussion}

\begin{figure}[t]
\begin{center}
\includegraphics[width=8cm]{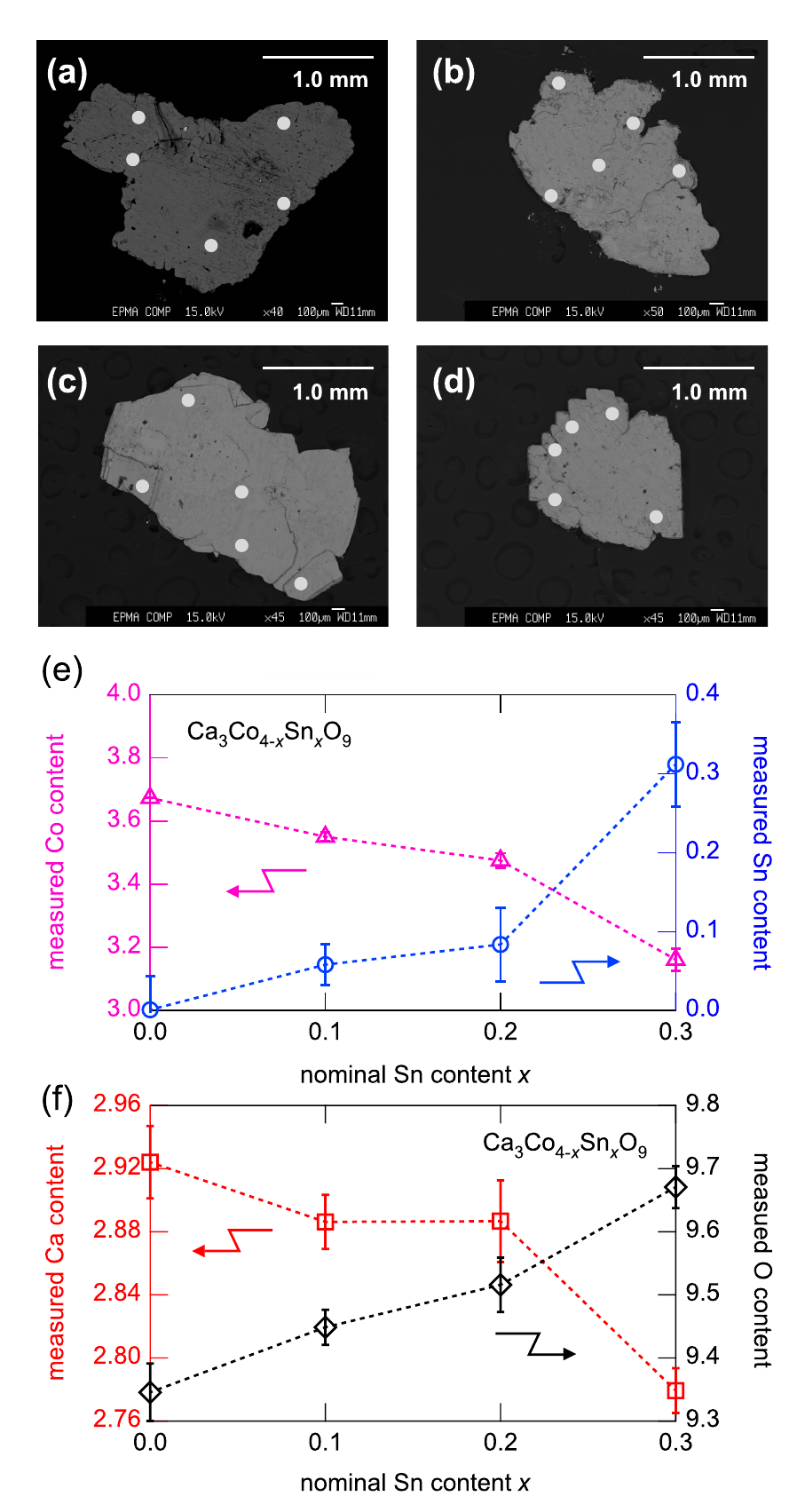}
\caption{
Composition images of Ca$_3$Co$_{4-x}$Sn$_x$O$_9$ single crystals
[(a) $x$ = 0, (b) $x$ = 0.1, (c) $x$ = 0.2, and (d) $x$ = 0.3] measured by the electron probe micro analyzer (EPMA) experiments. The measured points are marked with filled circles.
(e,f) Comparison of nominal and measured atomic contents in \sn single crystals ($0 \le x \le 0.3$) via EPMA measurements. 
}
\label{fig:epma}
\end{center}
\end{figure}

Figures \ref{fig:epma}(a)-(d) show the composition images of the \sn ($0 \le x \le 0.3$) single crystals.
Quantitative element analysis has been performed at various points marked with filled circles.
Figures \ref{fig:epma}(e) and \ref{fig:epma}(f) show a comparison of the nominal and measured atomic content in \sn single crystals.
The actual Sn content is slightly lower than expected, particularly for $x = 0.1$ and 0.2. The reason of this difference remains unknown at present, but as the single-crystal samples are prepared by the flux method, the compositional inhomogeneity may be present either between individual samples or spatially within a single sample.
As shown in \ref{fig:epma}(f), 
the Ca content has remained almost unchanged up to $x=0.2$, indicating that Sn replaces Co ions to induce electron doping.
On the other hand, the Ca content decreases in the $x=0.3$ sample, implying that Ca$^{2+}$ is partially replaced by Sn$^{2+}$ as suggested.
Nevertheless, the Co content indeed decreases in the $x=0.3$ sample as depicted in Fig. \ref{fig:epma}(e), indicating a successful Sn substitution into the Co sites.
In addition, 
considering that the ionic radius of Sn$^{2+}$ is 0.93 \AA~and that of Ca$^{2+}$ is 0.99 \AA~\cite{AHRENS1952155}, it is unlikely that Sn$^{2+}$ fully substitutes for Ca$^{2+}$, given the experimental observation that the lattice constants increase upon Sn substitution as discussed below.

\begin{figure}[t]
\begin{center}
\includegraphics[width=7cm]{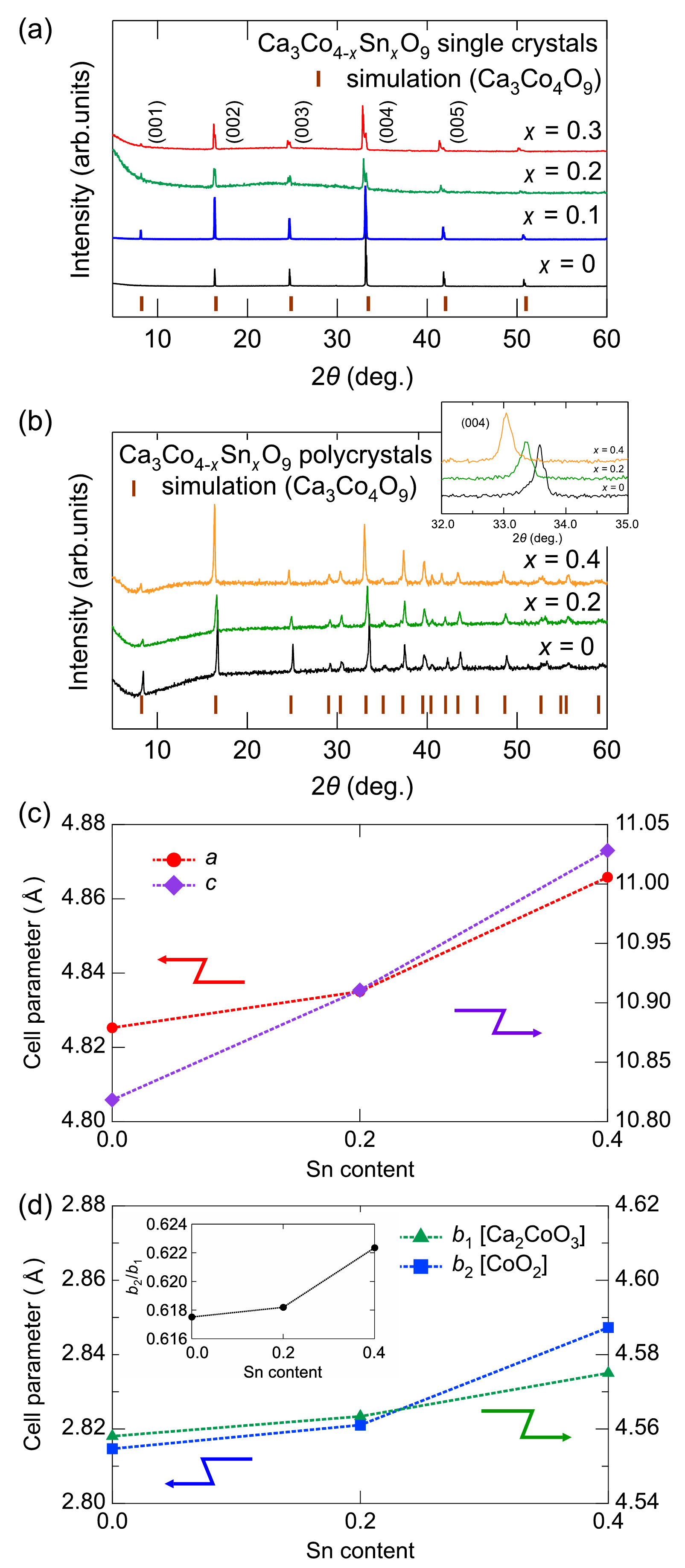}
\caption{
(a) X-ray diffraction (XRD) patterns of the as-grown single crystals of \sn ($0 \le x \le 0.3$) measured at room temperature.
(b) XRD patterns of the polycrystalline samples of \sn ($0 \le x \le 0.4$) measured at room temperature. 
Inset compares these data around the (004) reflection.
(c,d) Lattice parameters as a function of nominal Sn composition. 
$b_1$ and $b_2$ correspond to the $b$ parameters of [Ca$_2$CoO$_3$] and [CoO$_2$] layers, respectively. 
The inset of (d) shows the ratio $b_2/b_1$ of the parameter $b$ of the CoO$_2$ layer ($b_2$) to the parameter $b$ of the Ca$_2$CoO$_3$ layer ($b_1$).
}
\label{fig:xrd}
\end{center}
\end{figure}

The XRD patterns for single crystals of \sn ($0 \le x \le 0.3$) measured with the normal vector perpendicular to the $ab$-plane at room temperature are shown in Fig. \ref{fig:xrd}(a). 
Note that there is no impurity phase in the present results in contrast to the diffraction patterns of Bi-substituted \cco samples, in which an impurity phase is grown even in single crystals \cite{Mikami2006,Heieh2014}. 
We have also performed the powder XRD measurements using polycrystalline samples of \sn  ($x$ = 0, 0.2, 0.4) at room temperature
as shown in Fig. \ref{fig:xrd}(b). 
All data are assigned with the single phase and in agreement with the reported data for \cco structure. 
The inset of Fig. \ref{fig:xrd}(b) shows the diffraction peak of the (004) reflection, which systematically changes with the Sn substitution. 
Here, \cco consists of the rock-salt-type Ca$_2$CoO$_3$ layers and the CdI$_2$-type CoO$_2$ layers, with different values of $b_1$ and $b_2$ for the lattice parameters along the $b$-axis, respectively. 
The values of the lattice parameters estimated from the powder XRD measurements are shown in Figs. \ref{fig:xrd}(c) and \ref{fig:xrd}(d). 
It can be seen that the parameters $a$ and $c$ increase with increasing Sn substitution. 
The parameters $b$ in the Ca$_2$CoO$_3$ layer ($b_1$) and $b$ in the CoO$_2$ layer ($b_2$) also increase systematically with Sn substitution. 
This result suggests that Sn$^{4+}$ (0.69 \AA) substitutes low-spin Co$^{3+}$ (0.545 \AA),
resulting in the overall crystal expansion. 
Note that the change in the ratio $b_2/b_1$ for the substituted samples is crucial to determine the substitution sites \cite{Maignan2002,ChoJEM2015,Mohammed2020,Iqbal2023}.
The inset of Fig. \ref{fig:xrd}(d) shows the ratio $b_2/b_1$, 
which increases systematically with Sn substitution. 
This result implies that Sn substitution may exert a greater influence on the conductive CoO$_2$ layer with the lattice constant $b_2$ than the Ca$_2$CoO$_3$ layer with the lattice constant $b_1$,
the effect of which on the transport properties will be discussed later.

\begin{figure}[t]
\begin{center}
\includegraphics[width=8.5cm]{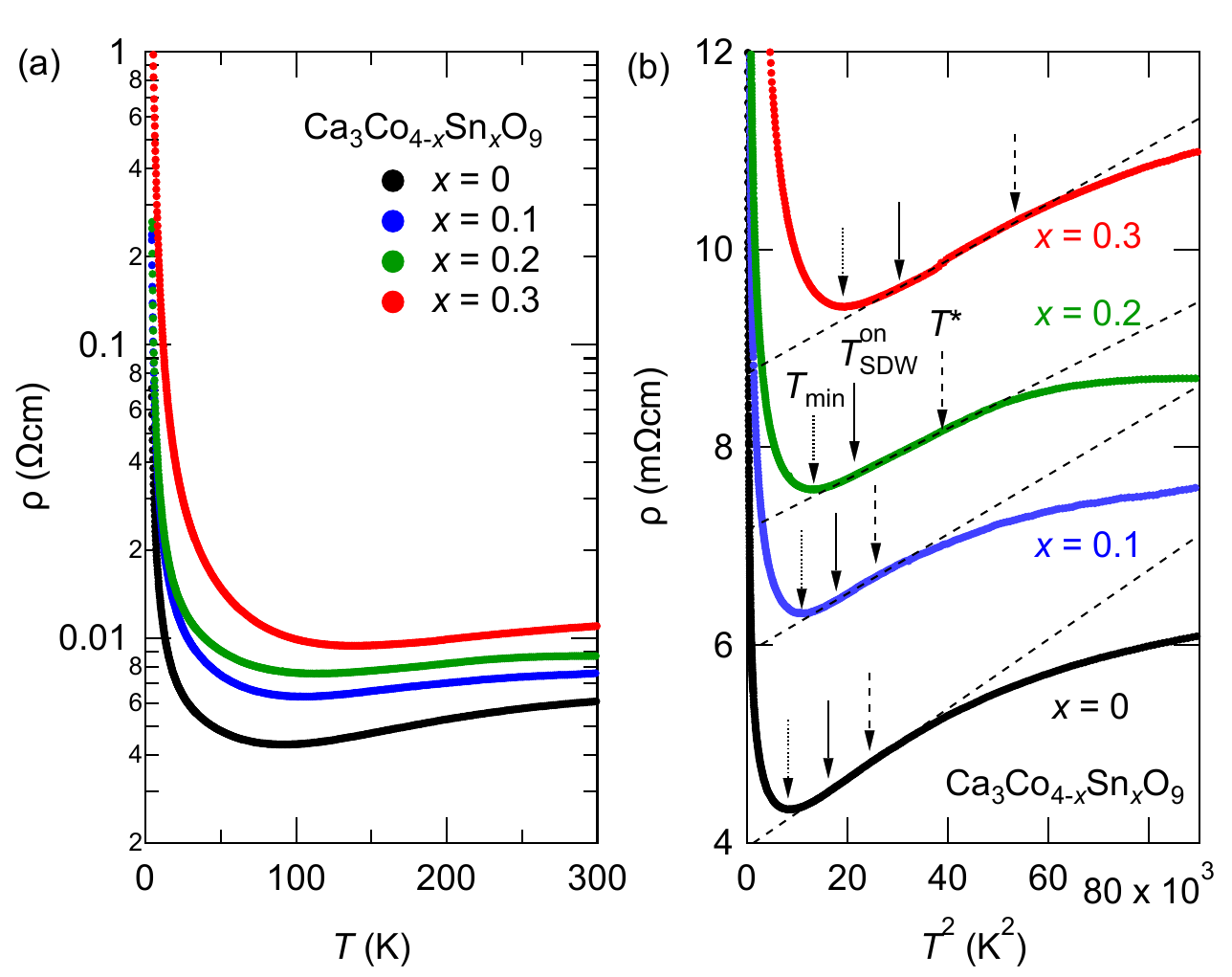}
\caption{(a) Temperature dependence of the resistivity $\rho$ for \sn ($0\le x \le 0.3$) single crystals in a semi-logarithmic plot. (b) The resistivity as a function of $T^2$. Dashed lines show the Fermi-liquid behavior of $\rho(T) = \rho_0+AT^2$, which is realized in the intermediate temperature range between the onset temperature $T^{\rm on}_{\rm SDW}$ (solid arrows) and the crossover temperature $T^*$ (dashed arrows).
A characteristic temperature $T_{\rm min}$, where the resistivity shows a minimum value, is also depicted as the dotted arrows.}
\label{fig:resistivity}
\end{center}
\end{figure}

The temperature dependence of the resistivity $\rho$ for \sn single crystals is shown in Fig. \ref{fig:resistivity}(a). 
All of the samples show similar behavior, but with increasing Sn content, the resistivity increases evidently. 
The increase of the electrical resistivity may be due to the decrease of the hole concentration and the increase of the scattering rate caused by
the tetravalent Sn$^{4+}$ substitution for the trivalent Co$^{3+}$ ions, which will be discussed later along with the results of the Seebeck coefficients.
In Fig. \ref{fig:resistivity}(b), we plot the resistivity as a function of $T^2$ for \sn single crystals.
As seen in earlier work~\cite{PhysRevB.71.233108}, 
the Fermi-liquid behavior of $\rho(T) = \rho_0+AT^2$ is observed 
in the intermediate temperature range between the SDW onset temperature $T^{\rm on}_{\rm SDW}$ and the crossover temperature $T^*$.
Note that, a characteristic temperature $T_{\rm min}$, where the resistivity shows a minimum value, can be interpreted as 
the onset temperature for the short-range SDW order \cite{Sugiyama2002}.
However, the SDW amplitude in the $\mu$SR data exhibits a gradual temperature dependence \cite{Sugiyama2002}, which seems to be difficult to precisely determine the onset temperature.
In addition, as discussed in Ref.~\cite{Heieh2014}, 
the carrier concentration decreases with decreasing temperature owing to the pseudo-gap opening in the SDW phase of \cco.
Therefore, the resistivity should deviate from the Fermi-liquid temperature variation below the SDW onset temperature, and it is thus reasonable to regard the lower temperature limit [depicted as solid arrows in Fig.\ref{fig:resistivity}(b)] of the Fermi-liquid regime as the onset temperature $T^{\rm on}_{\rm SDW}$.

\begin{figure}[t]
\begin{center}
\includegraphics[width=8cm]{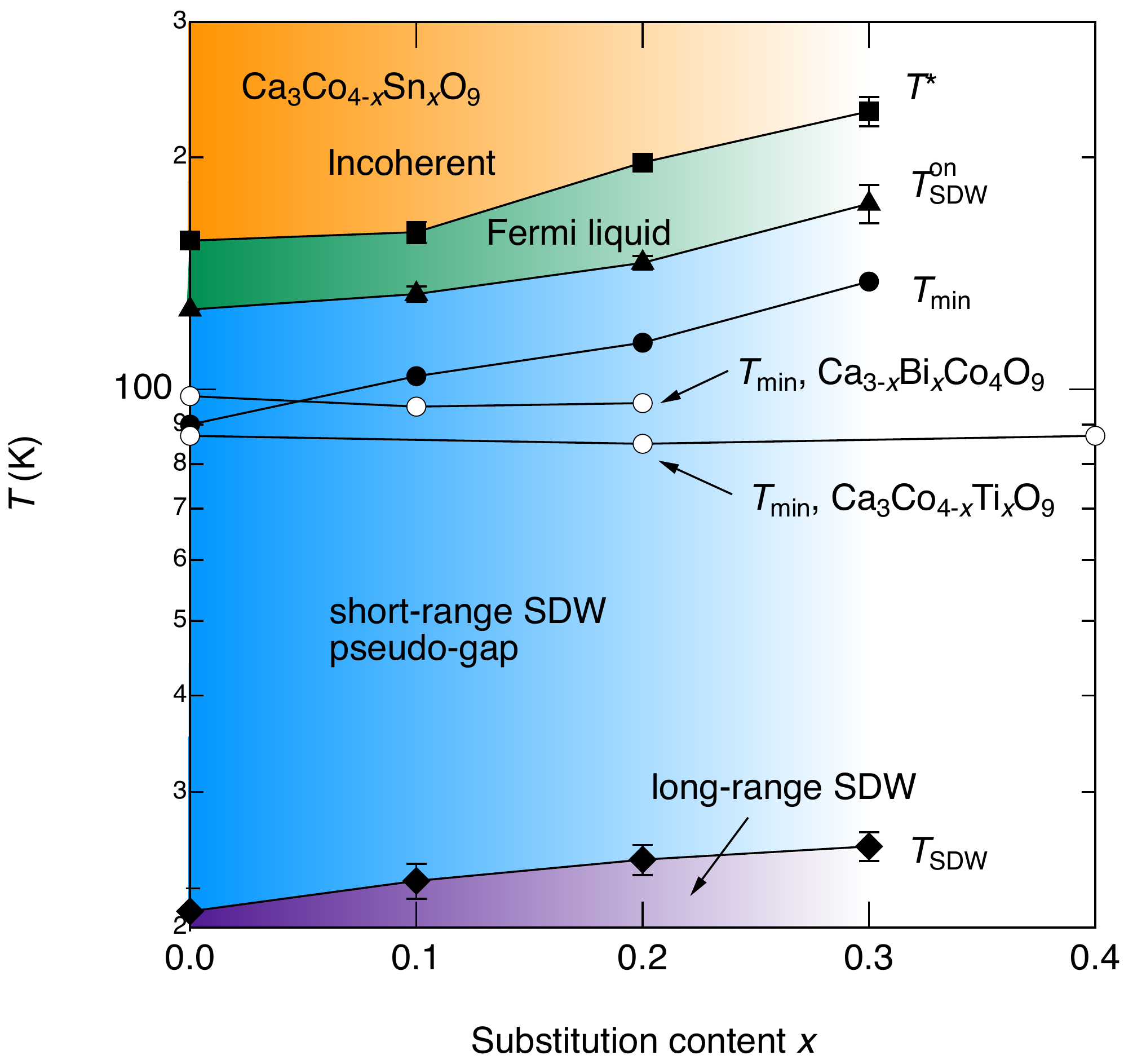}
\caption{
The characteristic temperatures as a function of the substitution content $x$ in \sn.
$T_{\rm min}$, $T_{\rm SDW}^{\rm on}$, $T^*$, and  $T_{\rm SDW}$ are plotted as the solid circles, triangles, squares, and diamonds respectively.
For comparison, $T_{\rm min}$ data in Ca$_3$Co$_{4-x}$Ti$_x$O$_9$ \cite{Zhao2006} and Ca$_{3-x}$Bi$_x$Co$_4$O$_9$ \cite{Heieh2014} are also plotted as the open circles.
}
\label{fig:phase}
\end{center}
\end{figure}

Nonetheless, it is clear that the SDW onset temperature increases with increasing Sn content $x$, indicating that
the SDW order becomes more stable with Sn substitution. 
Such an impurity-induced stabilization of the magnetic order has also been discussed in the layered cuprate superconductors \cite{PhysRevLett.99.147002,PhysRevLett.105.147002,Schmid_2013,JANG2023169164},
where the dynamical spin fluctuations are frozen out by the impurities.
It is also interesting to note that the charge density wave in the kagome lattice is also enhanced by substitutions \cite{PhysRevB.110.054518}.
Figure~\ref{fig:phase} shows the characteristic temperatures as a function of the substitution content $x$ in \sn along with other substituted systems \cite{Zhao2006,Heieh2014}.
In \sn, both  $T_{\rm min}$ and $T^{\rm on}_{\rm SDW}$ increase with increasing Sn content $x$, 
while $T_{\rm min}$ is essentially constant as a function of the atomic contents in the Ti- and Bi-substituted crystals.
As discussed previously,
the Sn ions mainly substitute the Co sites in the conductive layers as shown the inset of Fig.~\ref{fig:xrd}(d), indicating that the Sn substitution directly affects the SDW order.
In contrast, Bi and Ti ions tend to enter the block layers \cite{Zhao2006,Heieh2014}, which may less influence the SDW formation.
Thus, the observed difference in the content dependence of the characteristic temperatures among the substituted systems may originate from the site-dependent substitution effect in \cco.
We also note the effect from the oxygen content; a previous study has reported that an increase in the oxygen content induces a decrease in $T_{\rm min}$ \cite{Luo_2006}. On the other hand, as shown in Fig. \ref{fig:epma}(f), 
the oxygen content increases with increasing Sn content in the present study. 
Therefore, the present result that the increase in both $T_{\rm min}$ and $T^{\rm on}_{\rm SDW}$ occurs with increasing Sn content is not due to the oxygen non-stoichiometry.

\begin{figure}[t]
\begin{center}
\includegraphics[width=8cm]{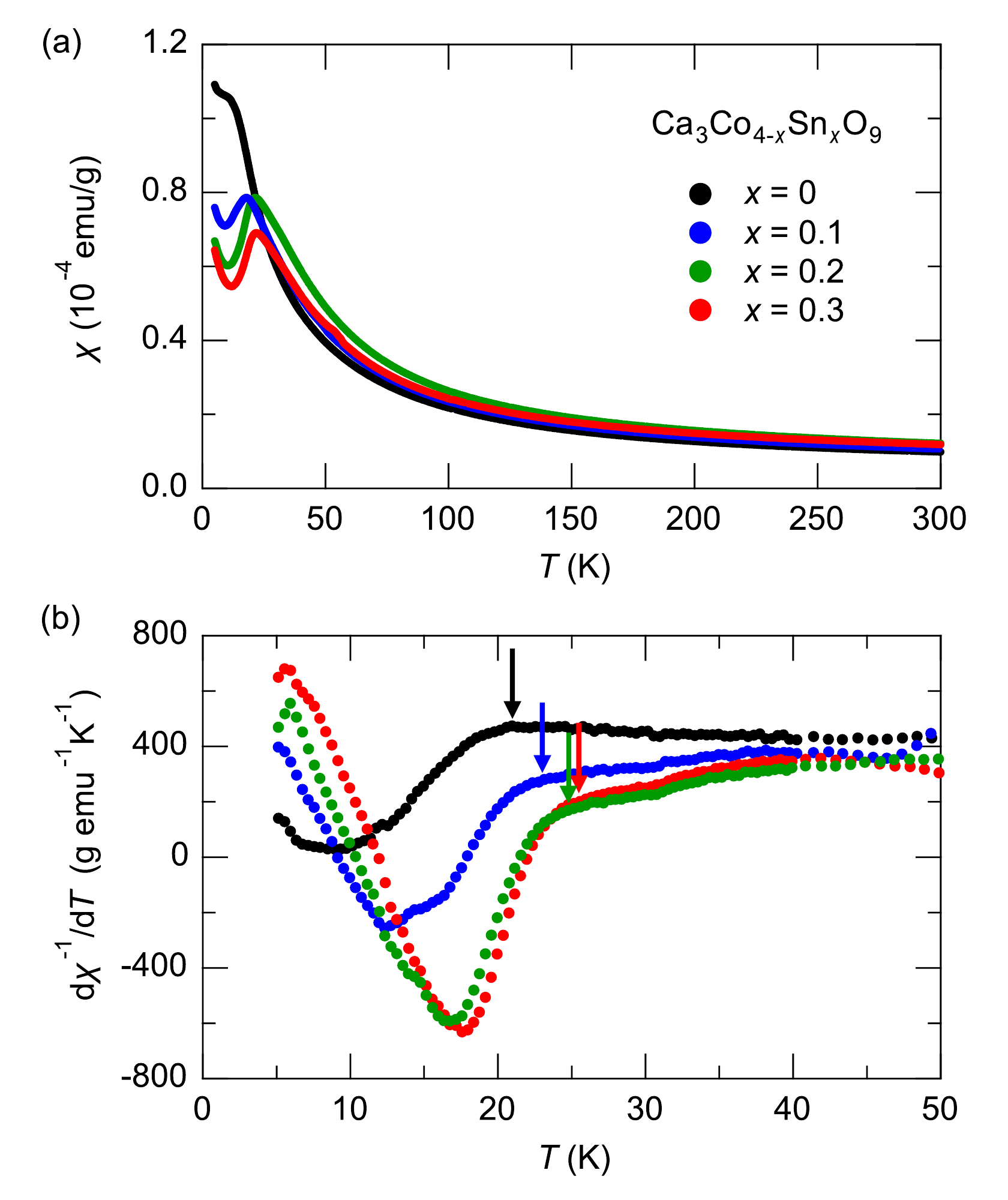}
\caption{(a) Temperature dependence of the magnetic susceptibility $\chi$ for \sn single crystals.
(b) Temperature dependence of $d\chi^{-1}/dT$ at low temperatures. 
The characteristic temperatures below which the high-temperature linear variations deviate are indicated by solid arrows.}
\label{fig:magchi}
\end{center}
\end{figure}

To examine the magnetic behavior of the present samples, we have also performed the magnetic susceptibility measurements.
In previous studies, the transition temperature for the long-range SDW order, $T_{\rm SDW}$, has been determined by several methods such as anomalies in $d\chi/dT$ \cite{Sugiyama2002,PhysRevB.68.134423}.
However, $\chi(T)$ shows Curie-like behavior of $\chi(T)\sim C/T$ in a whole temperature range as shown in Fig.~\ref{fig:magchi}(a), 
so one may obtain $T_{\rm SDW}$ as the deviation point from the high-temperature Curie-like behavior by using $d(\chi^{-1})/dT$ vs $T$ plot, as employed in Refs. \cite{Huang2013,Huang2013_2}.
In Fig.~\ref{fig:magchi}(b), we show the temperature dependence of $d(\chi^{-1})/dT$ for \sn single crystals, and the transition temperature increases with increasing Sn content. The Sn content dependence of $T_{\rm SDW}$ is also plotted in Fig.~\ref{fig:phase} and this tendency is consistent with the transport results.

\begin{figure}[t]
\begin{center}
\includegraphics[width=8cm]{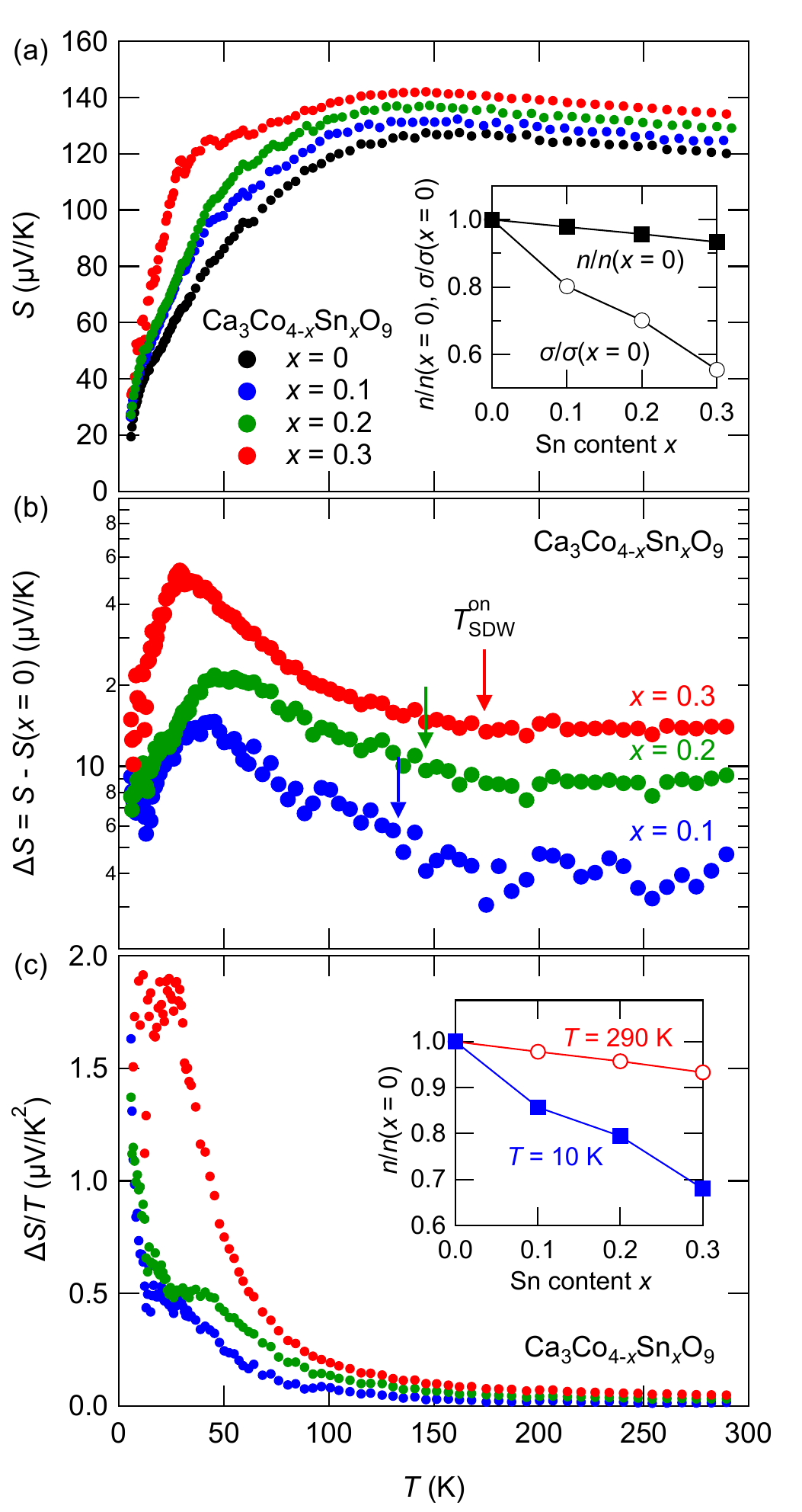}
\caption{(a) Temperature dependence of the Seebeck coefficient for \sn single crystals.
The inset shows the Sn content variations of the hole concentration estimated using the Heikes formula (solid squares)
and the electrical conductivity at $T=290$~K (open circles), both of which are normalized by the values of the undoped ($x=0$) sample.
(b) Temperature dependence of $\Delta S$ [$= S - S(x=0)$].
Solid arrows indicate the SDW onset temperatures obtained from the resistivity data in Fig.~\ref{fig:resistivity}(b).
(c) Temperature dependence of $\Delta S/T$. The inset shows the hole concentration estimated at $T=290$~K (open circles) and $T=10$~K (solid squares).}
\label{fig:seebeck}
\end{center}
\end{figure}

The enhancement of the SDW onset temperature and the associated pseudo-gap formation in \sn single crystals strongly affect the Seebeck coefficients.
Figure~\ref{fig:seebeck}(a) shows the temperature dependence of the Seebeck coefficient in \sn,
which systematically increases with increasing Sn content in the overall temperature range.
We now evaluate the hole concentration $n$ from the Seebeck coefficient at $T=290$~K using the modified Heikes formula
$S=-(k_{\rm B}/e)\ln[(g_3/g_4)n/(1-n)]$,
where
$k_{\rm B}$ is the Boltzmann constant, $e$ is the elementary charge, and
$g_3$ and $g_4$ are the spin-orbital degeneracy of Co$^{3+}$ and Co$^{4+}$ ions, respectively \cite{Koshibae2000,Koshibae2001}.
Note that both Co$^{3+}$ and Co$^{4+}$ are assumed to be in the low-spin state with $g_3=1$ and $g_4=6$ at room temperature \cite{Mizokawa2005,Klie2012}.
The hole concentration normalized by the value in the undoped sample $n/n(x=0)$ is then plotted as a function of the Sn content $x$ in the inset of Fig.~\ref{fig:seebeck}(a),
indicating a systematic electron doping effect with the Sn substitution.
Note that the scattering rate also increases with increasing Sn content, as seen in the 
Sn content dependence of the 
electrical conductivity measured at $T=290$~K normalized by the value in the undoped sample $\sigma/\sigma(x=0)$ [inset of Fig.~\ref{fig:resistivity}(a)].

Figure~\ref{fig:seebeck}(b) shows the temperature dependence of the difference between the Seebeck coefficients of the substituted samples and of the undoped sample, defined as $\Delta S = S - S(x=0)$.
Below the SDW onset temperature $T_{\rm SDW}^{\rm on}$, which is determined as the lower temperature limit of the Fermi-liquid regime shown in Fig.~\ref{fig:phase}, 
the hole concentration is further suppressed owing to the pseudo-gap opening, and
$\Delta S$ is then enhanced.
We also plot the temperature dependence of $\Delta S/T$ in Fig.~\ref{fig:seebeck}(c),
since the low-temperature Seebeck coefficient in a two-dimensional metal is expressed as 
\begin{align}
    S=
    \frac{\pi k_B^2}{2e\hbar c_0}\frac{m^*}{n}T
    \label{eq:lowS}
\end{align}
where $\hbar$ is the reduced Planck constant, $c_0$ is the $c$-axis length, and $m^*$ is the effective mass \cite{Kresin1990}.
The observed low-temperature increase of $\Delta S/T$ indicates that the hole concentration is more reduced at low temperatures and in highly substituted samples.
The inset of Fig.~\ref{fig:seebeck}(c) compares the normalized hole concentration at $T=10$~K obtained using Eq.~(\ref{eq:lowS}) with
the normalized hole concentration at $T=290$~K obtained using the Heikes formula.
Indeed, 
the hole concentration is suppressed more strongly at low temperatures and in highly substituted samples,
indicating that the SDW order and associated pseudo-gap formation become more robust with increasing Sn content.

Magnetic order caused by the non-magnetic impurities has been extensively investigated in the context of \textit{order by disorder} \cite{PhysRevLett.86.1086,PhysRevLett.103.047201}, where the nucleation of the magnetic order pattern occurs around a non-magnetic impurity. 
In fact, several correlated oxides with magnetic instabilities exhibit disorder-induced magnetic order, as is seen in ferromagnetism in Ca(Ru,$M$)O$_3$ ($M$ = Sc, Ti) \cite{PhysRevB.63.172403,Yamamoto2016} and incommensurate SDW order in Sr$_2$(Ru,Ti)O$_4$ \cite{PhysRevB.63.180504,PhysRevResearch.3.023067}.
Similarly, in the present compound \cco, the SDW order is enhanced by Sn substitution, indicating that Sn ions substituted in the conducting layers play a role in the acceleration of nucleation for the formation of the short-range SDW order, while it may be a future issue to theoretically examine the microscopic model of how magnetic ordering is induced by impurities.

\section{summary}
We have investigated the electrical and thermoelectric transport properties of the layered \sn single crystals.
The temperature dependence of the electrical resistivity exhibits similar trends in the undoped crystals, 
although it systematically increases with a higher Sn content $x$. 
Similarly, the Seebeck coefficient also increases with increasing Sn content. 
In the Sn-substituted samples, the SDW onset temperature significantly increases, probably owing to an impurity-induced stabilization of the SDW order as is also discussed in the layered cuprate superconductors.
Indeed, the magnetization measurements indicate that the transition temperature for the long-range SDW order also increases with increasing Sn content.
Such enhanced short-range (onset) and long-range transition temperatures may originate from the site-dependence substitution effect in \cco,
where the Sn ions tend to replace the Co ions in the conductive CoO$_2$ layers to strongly affect the SDW formation.

\section*{Acknowledgments}

We appreciate R. Kurihara and H. Yaguchi for discussion.
This work was supported by JSPS KAKENHI Grants No. 23K22437 and No. 24K06945.

\section*{DATA AVAILABILITY}

The data are available from the authors upon reasonable request.

\end{document}